\documentclass[
    ,final            % use final for the camera ready runs
  ]
  {aipproc}

\layoutstyle{6x9}
\begin{document}

\title{The hidden X-ray breaks in afterglow light curves}


%<Replace this text with PACS numbers; choose from this list:
%\texttt{http://www.aip.org/pacs/index.html}>}
%98.70.Rz 	¦Ã-ray sources; ¦Ã-ray bursts
%95.85.Nv 	X-ray observations
%95.55.Ka 	X- and ¦Ã-ray telescopes and instrumentation
%95.75.Pq 	Mathematical procedures and computer techniques


\author{P.A.~Curran  \thanks{e-mail: pcurran@science.uva.nl}}
{
  address={Astronomical Institute, University of Amsterdam, Kruislaan 403, 1098\,SJ Amsterdam, The Netherlands}
}

\author{A.J.~van~der~Horst}{
  address={University of Alabama, National Space Science and Technology Center, 320 Sparkman Drive, Huntsville, AL 35805, USA}
  ,altaddress={Astronomical Institute, University of Amsterdam, Kruislaan 403, 1098\,SJ Amsterdam, The Netherlands}
}

\author{R.A.M.J.~Wijers}{
  address={Astronomical Institute, University of Amsterdam, Kruislaan 403, 1098\,SJ Amsterdam, The Netherlands}
}

\author{R.L.C.~Starling}{
  address={Department~of~Physics~and~Astronomy, University~of~Leicester, University~Road, Leicester~LE1~7RH, UK}
}

\begin{abstract}

Gamma-Ray Burst (GRB) afterglow observations in the \emph{Swift} era have a perceived lack of achromatic jet breaks compared to the \emph{BeppoSAX}, or pre-\emph{Swift} era. Specifically, relatively few breaks, consistent with jet breaks, are observed in the X-ray light curves of these bursts. If these breaks are truly missing, it has serious consequences for the interpretation of GRB jet collimation and energy requirements, and the use of GRBs as standard candles. 

Here we address the issue of X-ray breaks which are possibly `hidden' and hence the light curves are misinterpreted as being single power-laws. We show how a number of precedents, including GRB\,990510 \& GRB\,060206, exist for such hidden breaks and how, even with the well sampled light curves of the \emph{Swift} era, these breaks may be left misidentified. We do so by synthesising X-ray light curves and finding general trends via Monte Carlo analysis. Furthermore, in light of these simulations, we discuss how to best identify achromatic breaks in afterglow light curves via multi-wavelength analysis. 
\end{abstract}

\maketitle

%%%%%%%%%%%%%%%%%%%%%%%%%%%%%%%%%%%%%%%%%%%%
%% MAINMATTER
%%%%%%%%%%%%%%%%%%%%%%%%%%%%%%%%%%%%%%%%%%%%

\section{Introduction}

Since the launch of the \emph{Swift} satellite, the standard picture of Gamma-Ray Burst (GRB) afterglows has been called into question by the lack of observed achromatic temporal breaks, up to weeks in a few bursts \cite{panaitescu2006:MNRAS369,burrows2007:astro.ph2633}. In some afterglows, a break is unobserved in both the X-ray and optical light curves, while in other bursts a break is observed in one regime but not in the other. 
In the \emph{BeppoSAX} era, most well sampled light curves were in the optical regime, while in the \emph{Swift} era most well sampled light curves are in the X-ray regime. Our expectations of the observable signature of a jet break are hence based on the breaks observed pre-\emph{Swift}, predominantly by optical telescopes, and the models which explained them \citep[e.g.][]{rhoads1997:ApJ487,rhoads1999:ApJ525,sari1999:ApJ519}. It is not clear that the breaks will be identical in both the X-ray and optical regimes.

\section{Hidden X-ray breaks}

%\subsubsection{GRB\,060206}

\begin{figure}
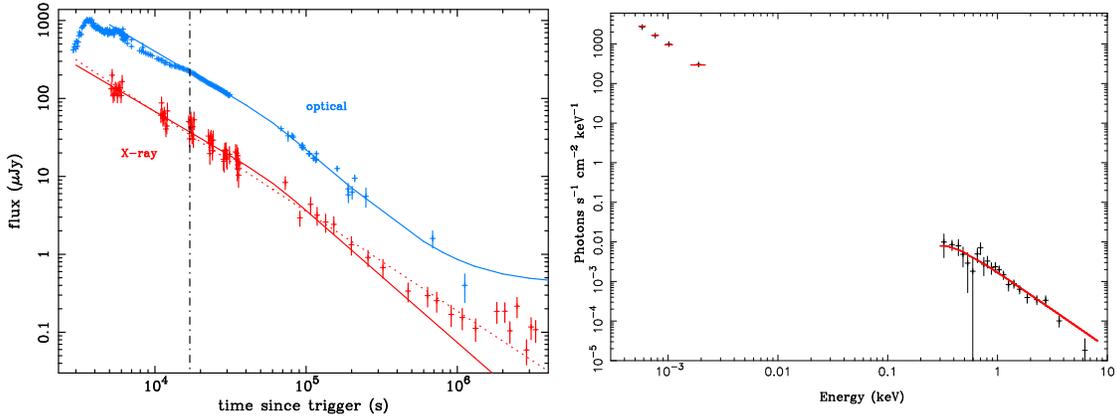
 \label{060206}
\hspace{-10mm}
\begin{minipage}{70mm}
 \includegraphics[height=.33\textheight,angle=270]{060206_lightcurve}
\end{minipage}
\hspace{3mm}
\begin{minipage}{65mm}
 \includegraphics[height=.33\textheight,angle=270]{060206_SED}
\end{minipage}
 \caption{\emph{Left} -- Optical ($R$-band, upper blue crosses) and X-ray count rate ($\times 200$, lower red crosses) light curves of GRB\,060206. The solid lines show the smoothly broken power-law (with host correction) fit to the optical data to the right of the vertical dot-dash line, and those same parameters scaled to the X-rays. The dotted line shows a single power-law fit to the X-rays. 
\emph{Right} -- The pre-break broad-band ($K_S$$H$$J$$R$, X-ray) SED of GRB\,060206 displays a single power-law spectral index. The post-break SED is consistent with this pre-break SED.}  
\end{figure}

We have previously shown \citep{curran2007:MNRAS381} that the well sampled X-ray afterglow of GRB\,060206 (Figure \ref{060206}, Left) can be described by a single power-law decay, though a broken power-law, which gives temporal indices and a break time similar to those in the optical, is as good a fit.
While it is difficult to accommodate the single power-law decay in the framework of the blast wave model \citep{zhang2004:IJMPA19}, an achromatic broken power-law decay can be interpreted in terms of a jet break or an energy injection break. 
The spectral indices of the optical to X-ray spectrum (Figure \ref{060206}, Right) are consistent before and after the optical break, i.e., at $\sim 2.9$ and $\sim 23$\,hours with the break at $\sim 16$\,hours ($\sim 3$\,hours in the rest frame) after the burst. This indicates that the temporal break is not caused by the passage of a break frequency through the optical regime in the broadband spectrum. 
The conclusion one can draw from this is that the break is caused by a change in the dynamics of the jet, e.g., the cessation of the energy injection phase or the beginning of the jet-spreading phase (the jet break interpretation).

%\subsubsection{GRB\,990510}

Many previously studied jet breaks do not display sharp changes in the temporal decay index, but a shallow roll-over from asymptotic values which is described by a smoothly broken power-law. 
The prototypical example of such a break is GRB\,990510 for which well sampled $B$, $V$, $R$ and $I$ band light curves display an achromatic break \citep[e.g.][]{stanek1999:ApJ522}. 
This is accepted as a jet break even though the X-ray light curve as measured by \emph{BeppoSAX} \citep{kuulkers2000:ApJ538} is satisfactorily described by a single power-law. 
A break at X-ray frequencies at the same time as the optical break is, however, not ruled out and the temporal slopes before and after that break are similar in the optical and X-rays. 
In the analysis of GRB\,060206 we are seeing the same phenomenon: the optical light curve displays a break, while the X-ray is satisfactorily described by a single power-law fit, though a broken power-law is not ruled out.
However, an X-ray break is necessary to explain the afterglow when interpreting it in the context of the blast wave model.

\section{Light curve synthesis}

\begin{figure} \label{monte_breaks}
\hspace{-10mm}
\begin{minipage}{70mm}
 \includegraphics[height=.33\textheight,angle=270]{broken}
\end{minipage}
\hspace{3mm}
\begin{minipage}{65mm}
\includegraphics[height=.24\textheight]{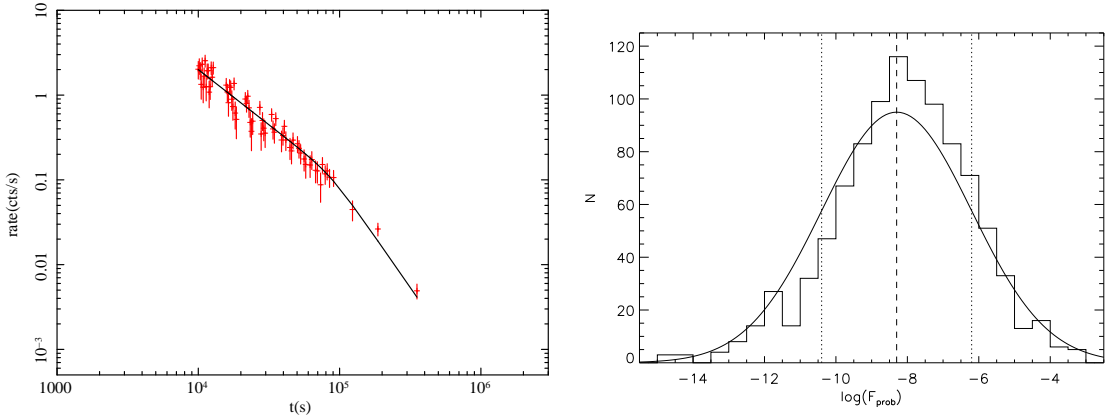}
\end{minipage}
 \caption{\emph{Left} --  A sample synthesised X-ray light curve including associated errors, overlaid on the underlying smoothly broken power-law with a break at 1\,day. 
\emph{Right} -- The distribution of the log of F-test probabilities for the light curve set described in the left panel, overlaid with a Gaussian. The lines show the average value and $1\sigma$ distribution. %, $\log(\rm{F_{prob}}) = -8.3 \pm 2.1$. 
}
\end{figure}

We have synthesised \emph{Swift} XRT light curves \citep{curran2007:arXiv0710} using similar criteria to those used for the XRT online repository \citep{evans2007:A&A469} so that the two would be easily comparable. The main points of the synthesis are that the light curves are normalised to 0.1\,cts/s at 1\,day and span from 3\,hours to 30 days unless a rate cut-off of $5 \times 10^{-4}$\,cts/s is reached before that. 
Each broken power-law light curve (Figure \ref{monte_breaks}, Left) is synthesised with given electron energy distribution index, break time, input smoothing factor, environment and spectral regime. We synthesised a number of light curves, each from different sets of these physical parameters that cover the observed ranges. 
Each synthesised light curve was fit with a single power-law and a broken power-law, and the improvement of the $\chi^2$ of the fits was estimated via the F-test. The data points of each light curve were perturbed and the fitting repeated multiple times in a Monte Carlo analysis (Figure \ref{monte_breaks}, Right). 

Our results show that, in some cases, the F-test does not favour the broken power-law over the single power-law (i.e., there is a high probability of a chance improvement of $\chi^2$), even though the underlying power-law is in fact broken. This leads, in a number of cases, to breaks being unidentified and we hence need to exercise caution in ruling out breaks with certainty. These hidden breaks are most prevalent in bursts with high levels of smoothing, or when the break is at late times ($>1$\,day). As the magnitude of the break is dependent on the environment and spectral regime, these also effect the misidentifications of breaks.

\vspace{-2.5mm}

\section{Conclusions}

We have shown that a well sampled X-ray light curve synthesised from an underlying broken power-law can often be satisfactorily fit by a single power-law decay, though a broken power-law is as good a fit. This would lead us, in the absence of further evidence, to assume a single power-law and rule out the possibility of a break. This effect is most pronounced in bursts with high levels of smoothing, as many previously studied jet breaks display. 
In these cases the breaks do not display sharp changes in the temporal decay index but a shallow roll-over from asymptotic values, well described by a smoothly broken power-law, which makes accurate determination of the temporal slopes and break time difficult.

This has significant implications for the analysis of the numerous X-ray light curves that the \emph{Swift} satellite has afforded us. For those X-ray light curves extending up to $\sim 1$\,day or longer, for which we do not have well sampled optical light curves, caution is required when making claims about the absence of breaks based solely on a comparison of the nominal fitted slopes. Multi-wavelength (optical \& X-ray) temporal and spectral data up to late times are required to make a confident statement on the absence or presence of a break, and to determine the break parameters.
This is particularly important when performing statistical analyses on a large sample of bursts, for making collimation corrected energy estimates, and for using GRBs as standard candles. 

Judging from the cases of GRB\,990510 and GRB\,060206, there may be a tendency, if not yet strongly significant, for X-ray light curves to have less pronounced or smoother breaks than optical light curves and more detailed theoretical models of jet breaks are necessary to clarify whether jet breaks could vary somewhat between wavebands.

%%%%%%%%%%%%%%%%%%%%%%%%%%%%%%%%%%%%%%%%%%%%%%%%
%% BACKMATTER
%%%%%%%%%%%%%%%%%%%%%%%%%%%%%%%%%%%%%%%%%%%%%%%%

\begin{theacknowledgments}
We thank A.P. Beardmore \& K.L. Page for useful discussions on the XRT.
PAC \& RAMJW gratefully acknowledge the support of NWO under grant 639.043.302, 
while RLCS gratefully acknowledges support from STFC.
%
%We acknowledge benefits from collaboration within the EU FP5 Research Training Network ``Gamma-Ray Bursts: An Enigma and a Tool" (HPRN-CT-2002-00294). 
%This work made use of data supplied by the UK Swift Science Data Centre at the University of Leicester and the High Energy Astrophysics Science Archive Research Center Online Service, provided by the NASA/GSFC.  
\end{theacknowledgments}

\bibliographystyle{aipproc}   % if natbib is available
\bibliography{sf_proc}

\begin{thebibliography}{11}
\expandafter\ifx\csname natexlab\endcsname\relax\def\natexlab#1{#1}\fi
\providecommand{\enquote}[1]{``#1''}
\expandafter\ifx\csname url\endcsname\relax
  \def\url#1{\texttt{#1}}\fi
\expandafter\ifx\csname urlprefix\endcsname\relax\def\urlprefix{URL }\fi
\providecommand{\eprint}[2][]{\url{#2}}

\bibitem[{Panaitescu} et~al.(2006)]{panaitescu2006:MNRAS369}
A.~{Panaitescu}, P.~{M{\'e}sz{\'a}ros}, D.~{Burrows}, J.~{Nousek},
  N.~{Gehrels}, P.~{O'Brien}, and R.~{Willingale}, \emph{{MNRAS}} \textbf{369},
  2059--2064 (2006), \eprint{arXiv:astro-ph/0604105}.

\bibitem[{Burrows} and {Racusin}(2007)]{burrows2007:astro.ph2633}
D.~N. {Burrows}, and J.~{Racusin}, \emph{Il Nuovo Cimento B} \textbf{121}, 1273
  (2007), \eprint{astro-ph/0702633}.

\bibitem[{Rhoads}(1997)]{rhoads1997:ApJ487}
J.~E. {Rhoads}, \emph{{ApJ}} \textbf{487}, L1+ (1997),
  \eprint{arXiv:astro-ph/9705163}.

\bibitem[{Rhoads}(1999)]{rhoads1999:ApJ525}
J.~E. {Rhoads}, \emph{{ApJ}} \textbf{525}, 737--749 (1999),
  \eprint{arXiv:astro-ph/9903399}.

\bibitem[{Sari} et~al.(1999)]{sari1999:ApJ519}
R.~{Sari}, T.~{Piran}, and J.~P. {Halpern}, \emph{{ApJ}} \textbf{519}, L17--L20
  (1999), \eprint{arXiv:astro-ph/9903339}.

\bibitem[{Curran} et~al.(2007{\natexlab{a}})]{curran2007:MNRAS381}
P.~A. {Curran}, A.~J. {van der Horst}, R.~A.~M.~J. {Wijers}, R.~L.~C.
  {Starling}, A.~J. {Castro-Tirado}, J.~P.~U. {Fynbo}, J.~{Gorosabel}, A.~S.
  {J{\"a}rvinen}, D.~{Malesani}, E.~{Rol}, N.~R. {Tanvir}, K.~{Wiersema}, M.~R.
  {Burleigh}, S.~L. {Casewell}, P.~D. {Dobbie}, S.~{Guziy}, P.~{Jakobsson},
  M.~{Jel{\'{\i}}nek}, P.~{Laursen}, A.~J. {Levan}, C.~G. {Mundell},
  J.~{N{\"a}r{\"a}nen}, and S.~{Piranomonte}, \emph{MNRAS} \textbf{381}, L65
  (2007{\natexlab{a}}), \eprint{arXiv:0706.1188}.

\bibitem[{Zhang} and {M{\'e}sz{\'a}ros}(2004)]{zhang2004:IJMPA19}
B.~{Zhang}, and P.~{M{\'e}sz{\'a}ros}, \emph{Int. J. of Mod. Phys. A}
  \textbf{19}, 2385--2472 (2004), \eprint{astro-ph/0311321}.

\bibitem[{Stanek} et~al.(1999)]{stanek1999:ApJ522}
K.~Z. {Stanek}, P.~M. {Garnavich}, J.~{Kaluzny}, W.~{Pych}, and I.~{Thompson},
  \emph{{ApJ}} \textbf{522}, L39--L42 (1999), \eprint{arXiv:astro-ph/9905304}.

\bibitem[{Kuulkers} et~al.(2000)]{kuulkers2000:ApJ538}
E.~{Kuulkers}, L.~A. {Antonelli}, L.~{Kuiper}, J.~S. {Kaastra}, L.~{Amati},
  E.~{Costa}, F.~{Frontera}, J.~{Heise}, J.~J.~M. {in 't Zand}, N.~{Masetti},
  L.~{Nicastro}, E.~{Pian}, L.~{Piro}, and P.~{Soffitta}, \emph{{ApJ}}
  \textbf{538}, 638--644 (2000), \eprint{arXiv:astro-ph/0003258}.

\bibitem[{Curran} et~al.(2007{\natexlab{b}})]{curran2007:arXiv0710}
P.~A. {Curran}, A.~J. {van der Horst}, and R.~A.~M.~J. {Wijers}, \emph{ArXiv
  e-prints} \textbf{710} (2007{\natexlab{b}}), \eprint{0710.5285}.

\bibitem[{Evans} et~al.(2007)]{evans2007:A&A469}
P.~A. {Evans}, A.~P. {Beardmore}, K.~L. {Page}, L.~G. {Tyler}, J.~P. {Osborne},
  M.~R. {Goad}, P.~T. {O'Brien}, L.~{Vetere}, J.~{Racusin}, D.~{Morris}, D.~N.
  {Burrows}, M.~{Capalbi}, M.~{Perri}, N.~{Gehrels}, and P.~{Romano},
  \emph{A\&A} \textbf{469}, 379--385 (2007), \eprint{arXiv:0704.0128}.

\end{thebibliography}

\end{document}